# Phonon transmission across $Mg_2Si/Mg_2Si_{1-x}Sn_x$ interfaces: A first-principles-based atomistic Green's function study

Xiaokun Gu,[1] Xiaobo Li,[2] and Ronggui Yang[1,3, a)]

[1]*Department of Mechanical Engineering, University of Colorado at Boulder, Colorado 80309, USA*

[2]*School of Energy and Power Engineering, Huazhong University of Science and Technology, Wuhan, Hubei 430074, People's Republic of China*

[3]*Material Science and Engineering Program, University of Colorado at Boulder, Colorado 80309, USA*

**Abstract**

Phonon transmission across interfaces of dissimilar materials has been studied intensively in the recent years by using atomistic simulation tools owing to its importance in determining the effective thermal conductivity of nanostructured materials. Atomistic Green's function method with interatomic force constants from the first-principles calculations has evolved to be a promising approach to study phonon transmission in many not well-studied material systems. However, the direct first-principles calculation for interatomic force constants becomes infeasible when the system involves atomic disorder. Mass approximation is usually used, but its validity has not been tested. In this paper, we employ the higher-order force constant model to extract harmonic force constants from the first-principles calculations, which originates from the virtual crystal approximation but considers the local force-field difference. As a feasibility

---

[a)] Email: Ronggui.Yang@Colorado.Edu

demonstration of the proposed method that integrates higher-order force constant model from the first-principles calculations with the atomistic Green's function, we study the phonon transmission in the $Mg_2Si/Mg_2Si_{1-x}Sn_x$ systems. When integrated with the atomistic Green's function, the widely-used mass approximation is found to overpredict phonon transmission across $Mg_2Si/Mg_2Sn$ interface. The difference can be attributed to the absence of local strain field-induced scattering in the mass approximation, which makes the high-frequency phonons less scattered. The frequency-dependent phonon transmission across an interface between a crystal and an alloy, which often appears in high efficiency "nanoparticle in alloy" thermoelectric materials, is studied. The interfacial thermal resistance across $Mg_2Si/Mg_2Si_{1-x}Sn_x$ interface is found to be weakly dependent on the composition of Sn when the composition x is less than 40%, but increases rapidly when it is larger than 40% due to the transition of high-frequency phonon DOS in $Mg_2Si_{1-x}Sn_x$ alloys. This work could have a great impact on the design of novel nanostructures with tunable thermal properties.

## I. Introduction

Understanding phonon transport in nanostructured materials and devices is of great importance in many technological fields, from thermal management in electronics and photonics[1, 2] to thermoelectric energy conversion,[3, 4] to thermal insulation and thermal protection system.[5] Interfaces play a critical role in determining phonon dynamics and thermal conduction in nanostructures. The detailed information of how phonon quanta with specific energy and momentum is scattered by an interface is essential for designing nanostructures with desirable thermal performance using mesoscopic modeling tools, such as phonon Boltzmann transport equation (BTE) based method and Monte Carlo simulations.[6-9]

In recent years, the atomistic Green's function (AGF) approach has been shown to be an efficient method to study frequency-dependent phonon transport across interfaces of dissimilar materials.[10, 11] The phonon transmission is highly dependent on the details of atomic configuration and interatomic interaction around the interface.[12, 13] AGF approach has been applied to study a wide range of interfaces, including sharp (smooth) interfaces,[14-16] rough interfaces, interfaces with vacancy defects and alloyed interfaces.[12] Empirical potentials are often used in AGF calculations to describe the interatomic interactions.[12-15, 17] The interatomic force constants from the first-principles calculations have also been integrated with AGF method for studying interfaces of materials, especially when the empirical potentials are not readily available.[16, 18, 19]

However, it would be quite challenging to extract the interatomic force constants of realistic material interfaces from the first-principles calculations. Comparing to the first-principles prediction of thermal conductivity of bulk crystals,[20-24] the lattice near an interface of two dissimilar materials is likely to be distorted due to the lattice mismatch and the difference in the force field experienced by the atoms in the interfacial region. A large supercell would be required for the first-principles calculations to capture the essential characteristics of the lattice-mismatched interface, which leads to severe numerical challenges. The mass approximation (MA)[16, 25] that was employed to calculate the thermal conductivity of alloys[22, 26, 27] has thus been used to extract interatomic force constants of the interfacial region from the first-principles. Under the MA, the differences in the lattice constants and the force fields between the two dissimilar materials are ignored and only the difference in atomic mass is taken into account. It was recently pointed out that the MA tends to overestimate the thermal conductivity due to the neglect of the local force-field difference.[28] Other studies showed that the MA under-predicts the contribution of high-frequency phonons, which leads to a lower thermal conductivity.[29] It is unclear whether MA is a reasonable approximation when integrated with the AGF approach to calculate the phonon transmission across the interfaces of dissimilar materials where there exist the differences in both lattice constants and force fields.

In this paper, we propose an integrated first-principles-based AGF approach using higher-order force constant model (HOFCM) to compute phonon transmission across interfaces of dissimilar materials. By considering the lattice distortion and the species-dependent local force field, the accuracy of the interatomic force constants extracted is improved compared with that from the MA, while the required computational resources are significantly less severe than directly extracting force constants from the standard first-principles calculations. As an example, we present the detailed studies of frequency-dependent phonon transmission across $Mg_2Si/Mg_2Si_{1-x}Sn_x$ interface, which is promising thermoelectric material system for intermediate temperature range energy conversion applications.[30-32] We systematically study how the lattice mismatch-induced local force field influences phonon transmission across the interfaces between dissimilar materials, which are inevitably ignored in the MA. This study could provide guidance for designing nanostructured materials with tunable thermal conductivity.

## II. Modeling Approaches

The AGF approach needs the harmonic interatomic force constants as inputs for the dynamical matrix to calculate phonon transmission. The conventional standard method that is used to extract the harmonic interatomic force constants of bulk crystals from the first-principles calculations is infeasible for interfaces where the lattice at the interfacial region is likely distorted due to the lattice mismatch and the local force fields. In this work, we employ the HOFCM, which originates from the MA but considers the local force difference due to the different species, to efficiently calculate the harmonic force constants of the interfacial regions, and then integrate the obtained force constants from the first-principles calculations with the AGF approach. In Sec. II A, we discuss the methods to extract the harmonic constants, including the standard direct methods from the first-principles, the MA, and the HOFCM. A brief summary of the AGF method is given in Sec. II B where the details can be found in literature including the author's prior work.[12, 15] The computational details are given in Sec. II C for $Mg_2Si/Mg_2Si_{1-x}Sn_x$ interface.

### A. Interatomic force constants from the first-principles calculations

The harmonic interatomic force constants $K$ that are used in AGF approach are the second-derivatives of the total energy of a system $E$ with respect to the atom displacements

$$K_{\mathbf{RR'}}^{\alpha\beta} = \frac{\partial^2 E}{\partial u_{\mathbf{R}}^{\alpha} \partial u_{\mathbf{R'}}^{\beta}}, \tag{1}$$

where $u_\mathbf{R}^\alpha$ denotes the displacement of an atom along the $\alpha$ direction whose equilibrium position is $\mathbf{R}$. In principle, one could compute the harmonic force constants of a given atomic system from the first-principles by monitoring the response of the total energy of the whole system under small perturbations, such as the small-displacement method[33-35] or the perturbation approach.[36] For a bulk crystal with a small primitive unit cell, the calculations for harmonic force constants using both methods are computationally affordable. Take the small-displacement method as an example, in which the force constants are calculated in a supercell. A series of first-principles calculations for total energy are performed when one atom in the primitive unit cell is moved with a small distance $\Delta$ in the order of 0.01 Å along a Cartesian direction. When all atoms in the primitive unit cell are displaced, the harmonic force constants are then extracted through either the finite difference scheme[33, 35] or fitting the displacement-force relation.[34] Although only the harmonic force constants involving the atoms in the primitive cell are obtained directly, other harmonic force constants between any two atoms in the crystal are extracted simultaneously by taking advantage of the periodicity of the crystal.

However, when dealing with interfaces, such conventional methods would become computationally overwhelming due to the loss of the periodicity of the crystal. The force constants between every atom pair need to be determined independently at least along the direction perpendicular to an interface. In addition, the distortion of lattice due to the mismatch of lattice constants could extend to several unit cells away from the interface, while the species mixing near the interface might span a few nanometers as well. As a result, very large supercells are needed for extracting the force constants. For example, if we use a supercell with a cross-sectional area of $2\times 2$ conventional unit cells to model an $Mg_2Si/Mg_2Sn$ interfacial region with species mixing that spans 10 unit cells, we have to perform a series of first-principles calculations to generate the displacement-force relation. In each calculation, one atom is displaced from their equilibrium position and the forces on other atoms are recorded. The total number of these first-principles calculations would be at least as large as $40 \times 12 \times 3 \times 2 = 2880$ (40 conventional unit cells, 12 atoms per unit cell for $Mg_2Si/Mg_2Sn$ system, 3 degrees of freedom, and 2 directions along a Cartesian direction). Furthermore, the calculations have to be repeated when the atomic configuration is slightly changed in order to obtain the harmonic force constants of the new system. For convenience, we refer the method just discussed above that the entire interfacial region is modeled to extract harmonic force constants as the direct method.

Due to the computational challenge in the direct method, the mass approximation that was employed to calculate the thermal conductivity of alloys is adopted recently, which models the interfacial region made up of two dissimilar materials to an imaginary (virtual) perfect homogenous bulk crystal so that the force constants can be easily extracted in first-principles calculations with the same procedures used for bulk crystals instead of modeling the entire interfacial region. In the density-functional-theory-based (DFT-based) approach, the type of atom is distinguished by its pseudopotential. A widely-used treatment of the MA in the DFT-based approach is to calculate the force constants in a virtual crystal,[37] where the two types of atoms in the simulation are replaced by virtual atoms, whose pseudopotential is the percentage weighted pseudopotentials of the two types of elements through

$$V_\sigma = \left[(1+\sigma)/2\right]V_{elem1}(\mathbf{r}) + \left[(1-\sigma)/2\right]V_{elem2}(\mathbf{r}), \tag{2}$$

where $V_{elem1}(\mathbf{r})$ and $V_{elem2}(\mathbf{r})$ are the pseudopotentials for element 1 and element 2, and $\sigma$ represents the likeliness of the virtual atom to be element 1 or 2. $\sigma=1$ and -1 represents element 1 and 2, respectively. When $\sigma=0$, the atom is of the averaged properties of element 1 and element 2. $\sigma=0$ is the most natural choice for the virtual atoms for an interface constructed by connecting two pure materials with element 1 and element 2.

Since the interfacial region can be treated as the homogenous crystal under the MA, the force field throughout the interfacial region is identical. As a result, the non-uniformity of the strain field resulted from the lattice mismatch in the interfacial region becomes absent. In addition, the local force fields within the bulk phases are no longer the actual ones, which should be determined from the direct method with the species' own pseudopotentials, but shifted to an averaged one. In other words, the vibrational properties of the reservoirs (semi-infinite crystal) could be inaccurate.

To overcome the limitations of the MA, we employ the HOFCM proposed by de Gironcoli[38] to extract the harmonic force constants for the interfacial region. The HOFCM was originally proposed to better describe the vibrational properties of Si/Ge systems beyond single crystals, including superlattices[39] and homogeneous $Si_{1-x}Ge_x$ alloys.[40] Taking advantage of the HOFCM, the first-principles-based calculations can reproduce the Raman spectra of $Si_xGe_{1-x}$ alloy very accurately.[38] In this work, we first approximate an interfacial region of large dimension that contains two species as a virtual crystal. Comparing the realistic interfacial region with the virtual crystal, the difference of their total energies originates from two aspects: 1) the atoms in a real interfacial region are not uniformly distributed as in a virtual crystal but are of small displacements $\mathbf{u_R}$ from their virtual crystal counterparts because the lattice is likely to be distorted near the interface. 2) The type of the atom $\sigma_\mathbf{R}$ and the corresponding force field in the real system is different from the virtual atom. Both of them make the harmonic force constants of the real interface deviated from the virtual crystal. The higher-order force constants with respect to $\mathbf{u_R}$ and $\sigma_R$ is then used to refine the total energy of the virtual crystal so that the difference between the real interfacial system and the virtual crystal is eliminated.

The difference between the total energy of the real interfacial region and that under the virtual crystal approximation is calculated using the Taylor's expansion of the total energy of the reference virtual crystal with respect to $\{\mathbf{u_R}\}$ and $\{\sigma_R\}$ as

$$\begin{aligned}
E(\{\mathbf{u_R}\},\{\sigma_R\}) &= E_{VC} + \sum_\mathbf{R}\left(\left.\frac{\partial E}{\partial \sigma_\mathbf{R}}\right|_{VC}\sigma_\mathbf{R} + \left.\frac{\partial E}{\partial \mathbf{u_R}}\right|_{VC}\cdot\mathbf{u_R}\right)\cdots \\
&= E_{VC} + \left(\sum_\mathbf{R}\left.\frac{\partial E}{\partial \sigma_\mathbf{R}}\right|_{VC}\sigma_\mathbf{R} + \sum_{\mathbf{RR'}}\left.\frac{\partial^2 E}{\partial \sigma_\mathbf{R}\partial \sigma_{\mathbf{R'}}}\right|_{VC}\sigma_\mathbf{R}\sigma_{\mathbf{R'}}+...\right) \\
&\quad + \left(\sum_\mathbf{R}\left.\frac{\partial E}{\partial \mathbf{u_R}}\right|_{VC}\cdot\mathbf{u_R} + \sum_{\mathbf{RR'}}\left.\frac{\partial^2 E}{\partial \mathbf{u_R}\partial \mathbf{u_{R'}}}\right|_{VC}\cdot\mathbf{u_R}\cdot\mathbf{u_{R'}} + \sum_{\mathbf{RR'}}\left.\frac{\partial^2 E}{\partial \mathbf{u_R}\partial \sigma_{\mathbf{R'}}}\right|_{VC}\cdot\mathbf{u_R}\cdot\sigma_{\mathbf{R'}}+...\right) \\
&= E_0 + \left(\sum_\mathbf{R}\left.\frac{\partial E}{\partial \mathbf{u_R}}\right|_{VC}\cdot\mathbf{u_R} + \sum_{\mathbf{RR'}}\left.\frac{\partial^2 E}{\partial \mathbf{u_R}\partial \mathbf{u_{R'}}}\right|_{VC}\cdot\mathbf{u_R}\cdot\mathbf{u_{R'}} + \sum_{\mathbf{RR'}}\left.\frac{\partial^2 E}{\partial \mathbf{u_R}\partial \sigma_{\mathbf{R'}}}\right|_{VC}\cdot\mathbf{u_R}\cdot\sigma_{\mathbf{R'}}+...\right)
\end{aligned}$$

(3),

where $E_{VC}$ is the total energy of the virtual crystal. Because the second and the third terms in the second line of Eq. (3) involve only the atom type $\{\sigma_R\}$, they do not influence the forces on the atoms but just shift the total energy. We can simply combine these terms with $E_{VC}$ and note as energy $E_0$, which is the total energy of the system taking into account atom type difference but with zero displacement. When the atom types are prefixed and the total energy is expanded up to the third-order terms with respect to the atom displacements, the expression can be further written as:

$$\begin{aligned}
E(\{\mathbf{u_R}\},\{\sigma_R\}) &= E_0 + \sum_{\mathbf{RR'}}\sum_\alpha G^\alpha_{\mathbf{RR'}}u^\alpha_\mathbf{R}\sigma_{\mathbf{R'}} + \frac{1}{2}\sum_{\mathbf{RR'}}\sum_{\alpha\beta}\phi^{\alpha\beta}_{\mathbf{RR'}}u^\alpha_\mathbf{R}u^\beta_{\mathbf{R'}} + \frac{1}{2}\sum_{\mathbf{RR'R''}}\sum_\alpha J^\alpha_{\mathbf{RR'R''}}u^\alpha_\mathbf{R}\sigma_{\mathbf{R'}}\sigma_{\mathbf{R''}} \\
&\quad + \frac{1}{2}\sum_{\mathbf{RR'R''}}\sum_{\alpha\beta}\chi^{\alpha\beta}_{\mathbf{RR'R''}}u^\alpha_\mathbf{R}u^\beta_{\mathbf{R'}}\sigma_{\mathbf{R''}} + \frac{1}{6}\sum_{\mathbf{RR'R''}}\sum_{\alpha\beta\gamma}\psi^{\alpha\beta\gamma}_{\mathbf{RR'R''}}u^\alpha_\mathbf{R}u^\beta_{\mathbf{R'}}u^\gamma_{\mathbf{R''}}
\end{aligned}$$

(4)

where the coefficients $\phi$ and $\psi$ are the derivatives of the total energy of the virtual crystal with respect to the displacement of atoms, which are the second-order harmonic and the third-order anharmonic force constants of the virtual crystal, and the coefficients $G$, $J$ and $\chi$ are the derivatives with respect to both atom displacements and atom types. Once these coefficients in Eq. (4) are determined, the difference of the total energy from a real interfacial system to that under the virtual crystal approximation can be greatly compensated for by using these terms.

To extract these coefficients in Eq. (4), we adopt the small-displacement method.[34] According to Eq. (4), the force on each atom is

$$F_\mathbf{R}^\alpha = -\frac{\partial E}{\partial \mathbf{u}_\mathbf{R}^\alpha} = -\sum_{\mathbf{R'}} G_{\mathbf{RR'}}^\alpha \sigma_{\mathbf{R'}} - \sum_{\mathbf{R'}}\sum_\beta \phi_{\mathbf{RR'}}^{\alpha\beta} \mathbf{u}_{\mathbf{R'}}^\beta - \frac{1}{2}\sum_{\mathbf{R'R''}} J_{\mathbf{R'R''}}^\alpha \sigma_{\mathbf{R'}} \sigma_{\mathbf{R''}}$$
$$-\sum_{\mathbf{R'R''}}\sum_\beta \chi_{\mathbf{RR'R''}}^{\alpha\beta} \mathbf{u}_{\mathbf{R'}}^\beta \sigma_{\mathbf{R''}} - \frac{1}{2}\sum_{\mathbf{R'R''}}\sum_{\beta\gamma} \psi_{\mathbf{RR'R''}}^{\alpha\beta\gamma} \mathbf{u}_{\mathbf{R'}}^\beta \mathbf{u}_{\mathbf{R''}}^\gamma$$
(5)

If the atoms in virtual crystal are displaced by a small distance away from their equilibrium positions or the types of atoms are altered, the force will not be zero anymore and could be accurately predicted in first-principles calculations through the Hellmann-Feynman theorem.[41] By fitting these displacement-type-force relations, the coefficients could be obtained.

We first determine the independent interatomic force constants based on the permutation symmetry of higher-order derivatives and the space group symmetry of the crystal.[34] Then one or two atoms in the virtual crystal are displaced from their equilibrium positions, and the types for all atoms are kept as $\{\sigma_\mathbf{R} = 0\}$. Through first-principles calculations, the forces on all atoms in the virtual crystal are computed. Putting the obtained forces, the atomic displacement and $\{\sigma_\mathbf{R} = 0\}$ into Eq. (5), we get a linear equation set with respect to $\phi$ and $\psi$ as unknown variables. Since the data set is usually larger than the number of independent force constants, the coefficients $\phi$ and $\psi$ cannot be solved exactly but are extracted from the linear fitting. To obtain physically correct interatomic force constants, the translational and rotational invariances[34, 38] of these interatomic force constants are further considered in fitting procedures and serve as constraints in the fitting procedures. We use the singular value decomposition (SVD) technique to convert the constrained linear least-squares problem to an unconstrained one. The details of the invariances of interatomic force constants and how to impose the invariances are discussed in Appendix. By solving the unconstrained equation sets, the coefficients $\phi$ and $\psi$ are extracted.

Since the coefficients $G$, $J$ and $\chi$ are related to both the atomic displacement and the atom types, we need to not only displace the atoms but also change the type of one or two atoms from 0 to $\pm 1$ so that the terms on $G$, $J$ and $\chi$ could take effects on the forces on the atoms in the system. Plugging the atomic displacements, the atom type $\{\sigma_\mathbf{R}\}$ and the values of the coefficients $\phi$ and $\psi$, which have been obtained before, Eq. (5) becomes the linear equation set with respect to $G$, $J$ and $\chi$. Similar to the coefficients $\phi$ and $\psi$, the physically correct coefficients $G$, $J$ and $\chi$ also need to satisfy their translational and rotational invariance relations,[38] which can be found in Appendix. With the SVD techniques, the coefficients $G$, $J$ and $\chi$ are solved from the linear equation set with the same least-squares method as we used to extract the coefficients $\phi$, $\psi$.

In AGF calculations, the atoms in the system need to be in the equilibrium positions. To determine the equilibrium configuration of the system, the analytical expression Eq. (4) are used to determine the equilibrium configuration of the system by minimizing the total energy of the system. The corresponding harmonic interatomic force constants $K$ used in AGF simulations are then given by taking the derivative of Eq. (4):

$$K^{\alpha\beta}_{\mathbf{RR'}} = \phi^{\alpha\beta}_{\mathbf{RR'}} + \sum_{\mathbf{R''}} \chi^{\alpha\beta}_{\mathbf{RR'R''}} \sigma_{\mathbf{R''}} + \sum_{\mathbf{R''}} \sum_{\gamma} \psi^{\alpha\beta\gamma}_{\mathbf{RR'R''}} \mathbf{u}^{\gamma}_{\mathbf{R''}}$$
(6)

There are a few advantages of using HOFCM for extracting the harmonic interatomic force constants compared with the direct method: 1) It avoids using a large supercell to model the interfacial region. Since all coefficients in Eq. (4) are the derivatives of the total energy of the virtual crystal, these coefficients can be extracted using a relatively small supercell just as that used to calculate the harmonic force constants in a bulk crystal with periodicity. 2) When the set of coefficients is obtained, it can be applied to systems with different atomic configurations. In contrast, in the direct method the force constants need to be recalculated every time when the configurations of interfacial regions are changed. 3) Using HOFCM, the system can be relaxed by minimizing the analytical expression of total energy, while system relaxation in DFT calculation requires a series of time-consuming self-consistent field calculations.

B. AGF approach for phonon transmission

A typical AGF simulation system for phonon transmission consists of three regions, one interfacial region and two semi-infinite reservoirs made up of bulk crystal 1 and 2. Phonon transmission across the interfacial region is calculated using the Green's function, which gives the response of the system under small perturbation. Under harmonic approximation, the Green's function $\mathbf{G}_{d,d}$ corresponding to the interfacial region can be calculated as,[11]

$$\mathbf{G}_{d,d} = \left[\omega^2 \mathbf{I} - \mathbf{H}_{d,d} - \Sigma_1 - \Sigma_2\right]^{-1},$$
(6)

where $\omega$ is the phonon frequency; $\mathbf{H}_{d,d}$ represents the dynamical matrix of the whole interfacial region, $\Sigma_1$ and $\Sigma_2$ is the self-energy matrices of the left and right reservoirs, which are calculated from the dynamical matrices of reservoirs.[14] The elements in the dynamical matrices are written as

$$H_{ij} = \frac{1}{\sqrt{m_\mathbf{R} m_{\mathbf{R}'}}} K_{\mathbf{R}\mathbf{R}'}^{ab}, \tag{7}$$

where $i$ ($j$) stands for the $i$th ($j$th) degree of freedom in the system, or the $\alpha$ ($\beta$) direction of the atom at $\mathbf{R}$ ($\mathbf{R}'$); $m_\mathbf{R}$ is the mass of the atom at $\mathbf{R}$. With the Green's function, the total phonon transmission across the interfacial region is calculated as,

$$\Xi(\omega) = Tr[\mathbf{\Gamma}_1 \mathbf{G}_{d,d} \mathbf{\Gamma}_2 \mathbf{G}_{d,d}^+], \tag{8}$$

where $\mathbf{\Gamma}_1 = i(\mathbf{\Sigma}_1 - \mathbf{\Sigma}_1^+)$, $\mathbf{\Gamma}_2 = i(\mathbf{\Sigma}_2 - \mathbf{\Sigma}_2^+)$ and " + " denotes the conjugate transpose of the matrix. We use transmission per phonon $\xi(\omega)$ to present our results, which is related to the total phonon transmission through,[15]

$$\Xi(\omega) = \xi(\omega) M(\omega), \tag{9}$$

where $M(\omega)$ is the total number of phonon modes at frequency ω from materials 1. More details can be found in Ref. [10, 12, 14].

The interfacial thermal conductance can then be calculated with the Landauer formalism using the total phonon transmission:[42]

$$G_0 = \frac{1}{2\pi A} \int_0^\infty \hbar\omega \frac{\partial f(\omega, T)}{\partial T} \Xi(\omega) d\omega, \tag{10}$$

where $A$ is the cross-sectional area perpendicular to the heat flow direction, $f(\omega, T)$ is the phonon occupation number at the temperature $T$. It has been recognized that Eq. (10) gives a finite conductance for phonon transport across a non-existing interface within a bulk material, which is different from the conventional definition of interfacial conductance.[43, 44] To eliminate the finite conductance for a non-existing interface, we adopt the revised expression used by Tian et al.[16] to calculate the thermal conductance,

$$G = \frac{G_0}{1-(G_0/G_1+G_0/G_2)/2}, \qquad (11)$$

where $G_1$ and $G_2$ are the thermal conductance of pure material 1 and pure material 2 calculated from Eq. (4).

C. Implementation for $Mg_2Si/Mg_2Si_xSn_{1-x}$ interface

To illustrate the feasibility of our proposed first-principles-based AGF approach with HOFCM, we implemented the calculation for $Mg_2Si/Mg_2Si_xSn_{1-x}$ thermoelectric material system. $Mg_2Si/Mg_2Sn$-based materials have recently been studied intensively as promising middle temperature-range (400-800 K) thermoelectric materials[31, 32, 45] due to their reasonably high thermoelectric figure of merit, ZT ~1.1 at 700 K, as well as the abundance of the constituent elements and their nontoxicity. Recent work has suggested that the thermoelectric performance of $Mg_2Si$-based materials could be further enhanced through nanostructuring, where interface scattering can significantly reduce the phononic thermal conductivity.[22] The understanding of phonon transmission and interfacial thermal conductance across $Mg_2Si/Mg_2Si_xSn_{1-x}$ interface would thus greatly benefit the design of high efficiency $Mg_2Si$-based thermoelectric materials with low thermal conductivity.

$Mg_2Si$ and $Mg_2Sn$ have cubic anti-fluorite (Fm-3m) structures:[46] Si and Sn atoms sit in the FCC sites while Mg atoms occupy the tetrahedral holes. The similarities and differences between $Mg_2Si$ and $Mg_2Sn$ are attributed to the nature of Si and Sn elements that sit in the same column of the periodic table. To apply the HOFCM, the Si and Sn atoms are replaced by a virtual element, whose pseudopotential is generated by averaging the pseudopotentials of Si and Sn elements according to Eq. (2). The generation of the pseudopotential of the virtual element has been implemented in plane wave package QUANTUM ESPRESSO.[47] Using the same package, DFT-based first-principles calculations are performed to generate the displacement-type-force relations, which are used to extract the interatomic force constants. Local density approximation (LDA) of Perdew and Zunger[48] with a cutoff energy of 40 Ry is used for the plane-wave expansion. Considering that the decay of the third-order interatomic force constants ($J, \psi, \chi$) is faster than that of the second-order ones ($G, \phi$) as the displacement distance between atoms is increased, we choose a smaller interaction cutoff for the calculation of the third-order interatomic force constants ($0.866a$) than that for the second-order ones ($1.5a$), where $a$ is the lattice constant. The choice of the cutoff for the third-order force constants is validated by the recent thermal conductivity calculations on $Mg_2Si$ and $Mg_2Sn$ based on Peierls-Boltzmann transport equation (PBTE) theory with interatomic force constants from DFT as inputs, where the same cutoff was applied.[22] A supercell made up of $3\times3\times3$ conventional unit cells with a $4\times4\times4$ Monkhorst-Pack mesh is employed to generate the displacement-type-force data needed for second-order force constants, where one atom is displaced or changed, while a smaller $2\times2\times2$ supercell with a denser $6\times6\times6$ Monkhorst-Pack mesh is used for the third-order force constants. All the coefficients that are needed in the HOFCM are extracted through the DFT calculations using the procedures discussed in section II.A. All the simulations are conducted in the virtual crystal with lattice constant, $a_0 = \left(a_0^{Mg_2Si} + a_0^{Mg_2Sn}\right)/2$, where $a_0^{Mg_2Si}$ and $a_0^{Mg_2Sn}$ are the lattice constants of $Mg_2Si$ and $Mg_2Sn$, respectively.

After the interatomic force constants are determined, the atomic system for the AGF simulation system is constructed and then the position of each atom is optimized until the force on each atom becomes zero. The optimization is performed using the fast inertial relaxation engine (FIRE) algorithm,[49] which is a molecular-dynamics-based method to find the minimum energy of an atomic system. The atomic positions are updated according to the forces that exerting on the atoms. The time steps and the atomic velocities are adjusted based on the forces and the velocities of the atoms so that the FIRE algorithm can efficiently relax the interfacial systems. The corresponding harmonic force constants $K$ of a given system in equilibrium are then calculated using Eq. (6).

In AGF simulations, when dealing with large system with thousands of atoms, large-size matrix operation is computationally costly. We employ a recursive AGF approach to efficiently calculate phonon transmission across the interfacial region, as done in our previous work [12, 15]

## III. Results and Discussion

To test the accuracy of the harmonic force constants from the HOFCM, the phonon dispersions of a few typical $Mg_2Si/Mg_2Sn$ structures are computed and compared with direct method in Sec. III(A). Phonon transmission across a sharp $Mg_2Si/Mg_2Sn$ interface is presented in Sec. III(B), while the phonon transmission and scattering across $Mg_2Si/Mg_2Si_{1-x}Sn_x$ interface are studied in Sec. III(C).

### A. Phonon dispersion

Since the lattice in the interfacial region is distorted and the atoms near the interface interact with other species, an accurate force field that can not only correctly describe the interatomic interaction within the unstrained bulk phases, but also the strained ones with specie mixing is essential. To evaluate the performance of the MA and the HOFCM on predicting the harmonic force constants in $Mg_2Si/Mg_2Sn$ systems, we study the phonon dispersion of the unstrained and strained $Mg_2Si$ and $Mg_2Sn$ crystals, and the phonon density of states (DOS) of a short-period $Mg_2Si/Mg_2Sn$ superlattice using the three first-principles-based methods to generate harmonic force constants: (a) the direct method with the actual pseudopentials of Si and Sn, (b) the MA with the pseudopotential generated according to Eq. (2) with $\{\sigma_\mathbf{R}\}=0$, and (c) the HOFCM, which improves the harmonic force constants from the MA with the higher-order terms, as detailed in Section II.(A).

Figure 1(a) shows the calculated phonon dispersion curves of $Mg_2Si$ and $Mg_2Sn$ in comparison with the neutron scattering experimental data.[50, 51] The overall agreement between the experimental data and the direct DFT calculations with Si/Sn pseudopotentials is excellent, except that a small difference (approximately 10% at Γ point) exhibits in the optical phonon modes. This small discrepancy is mainly due to the well-known red shift of the longitudinal optical (LO) phonon modes away from the transverse optical (TO) phonon modes (LO-TO splitting) in polar materials.[52] Since we do not consider the long-range interaction raised by the Born effective charge, the LO and TO modes intersect at the Γ point. While such a discrepancy in dispersion curves could be corrected by introducing a non-analytical term that is related to the Born effective charge into the dynamical matrix, taking all force constants in real space into account could be challenging. Since only a small portion of phonons is affected compared with the whole first Brillouin zone, we can expect that the LO-TO splitting has negligible effects on phonon transmission and the interfacial thermal conductance. To directly evaluate the influence of LO-TO on phonon transport, we also calculated phonon DOS with and without LO-TO splitting by including the non-analytical term into the dynamical matrix. Figure 1(b) shows that the DOS is almost unchanged by including LO-TO splitting.

The phonon dispersion curves of both $Mg_2Si$ and $Mg_2Sn$ calculated using the HOFCM almost overlap with that from the direct method, indicating that the HOFCM has the similar accuracy as that of the DFT calculations using the species' own pseudopotentials. It is rather clear that the MA underestimates the frequencies of both acoustic and optical modes of $Mg_2Si$, but overestimates those of $Mg_2Sn$. By examining the spring constants between Mg and its first-nearest neighbor Si (or Sn), defined as the trace of the harmonic force constants of Mg and Si(Sn) pairs, the bonding in $Mg_2Si$ is found to be 16% stiffer than $Mg_2Sn$. The bonding in the virtual crystal lies in the middle of those of $Mg_2Si$ and $Mg_2Sn$. Consequently, under the MA, the bonding stiffness in $Mg_2Si$ is weakened which leads to the downshift of phonon dispersion, while the opposite trend occurs in $Mg_2Sn$.

Figure 2 shows the phonon dispersion of the strained $Mg_2Si$ and strained $Mg_2Sn$, which are stretched or compressed to the averaged lattice constant of them to mimic the deformed lattice near the interface. Similar to the unstrained cases, the phonon dispersions calculated using the HOFCM are in good agreement with the direct method, but the results from the MA deviate to some extent. It is noted that the third-order anharmonic force constants do not participate in the refining of the harmonic force constants in the HOFCM since the relative positions of atoms in the strained crystal are unchanged compared with the reference structure where the force constants are extracted. The differences of the harmonic interatomic force constants for the strained cases examined here from the MA and the HOFCM are thus purely originated from the third-order force constants involving the atom type $\{\sigma_R\}$. This observation indicates the importance of taking the atom type into consideration when determining the accurate interatomic force constants, which is ignored in the conventional MA.

To test whether the HOFCM can represent the interactions between different species, we study the phonon properties of a superlattice, whose crystal structure can be regarded as a conventional unit cell of $Mg_2Si$ but with half of the Si atoms replaced by the Sn atoms (See. Fig. 3(a)). In this crystal, the Si atoms interact with the Sn atoms in a fashion similar to the interfacial region. The phonon DOS of this superlattice is shown in Fig. 3(b). While the curve of the phonon DOS from the HOFCM almost follows that from the direct method, there is considerable deviation for middle-to-high-frequency phonons in the MA from the direct method. The deviation could be related to the lattice deformation within the lattice. The phonons whose wavelengths are smaller than or comparable to the scale of lattice deformation are likely to be strongly affected.[53] Since the MA does not consider the lattice deformation, the transmission of the short wavelength phonons, which are usually middle-to-high-frequency phonons, are not accurately calculated using the MA.

B. Phonon transmission across a sharp $Mg_2Si/Mg_2Sn$ interface

To show the importance of accurate force constants on phonon transmission calculation, we study the phonon transmission across a sharp $Mg_2Si/Mg_2Sn$ interface. Two bulk $Mg_2Si$ and $Mg_2Sn$ crystals are connected with each other in a periodic box to form two interfaces, as shown in Fig. 4(a). The total length of the simulation box is 36 times of the average lattice constant $a_0$ of $Mg_2Si$ and $Mg_2Sn$, which is sufficient long to ensure that the atomic arrangement near the interface is independent to the choice of a larger length. After performing the minimization of the total energy of the crystal, we take half of the system with one interface as the interfacial region to conduct the AGF calculation, as shown in Fig. 4(b). The harmonic force constants of the interfacial region are generated according to atomic configuration after energy minimization.

The atoms near the middle region of each phase are evenly distributed due to the reflection symmetry, and these regions are used to calculate the harmonic force constants of the two reservoirs. For MA, we use Eq. (2) to generate the pseudopotential with $\{\sigma_\mathbf{R}\}=0$ for Mg$_2$Si/Mg$_2$Sn interfaces.

The phonon transmission curves from Mg$_2$Si to Mg$_2$Sn calculated from both the MA and the HOFCM are shown in Fig. 5(a). The phonon DOS of the two reservoirs are also presented in Fig. 5(b) and (c). Both phonon transmission curves show a similar trend. In the middle frequency region ($0.25-0.3\times10^{14}$ rad/s), there is a region with zero-transmission from both the HOFCM and the MA calculations. The zero phonon transmission is due to the frequency-gap between the acoustic and optical branches of Mg$_2$Sn. Any phonons in Mg$_2$Si with frequency lying in the gap of Mg$_2$Sn phonon dispersion cannot transmit, since Mg$_2$Sn cannot support any phonon modes with such frequencies. At the low frequency region below the gap, the phonon transmission is relatively high, close to unity. This is because the phonon DOS in Mg$_2$Sn side is much larger than that in Mg$_2$Si and there can be multiple phonons in the Mg$_2$Sn side with similar low-frequency to match the incoming phonons from the Mg$_2$Si side.[15] At the high frequency region, the transmissions for both cases are much lower than unity, indicating that high-frequency phonons are more likely to be scattered at the Mg$_2$Si/Mg$_2$Sn interface since the phonon dispersions of Mg$_2$Si and Mg$_2$Sn become quite different.

However, there are some quantitative differences. The MA predicts a smaller zero-transmission gap than HOFCM. The difference in the zero-transmission gap could be explained by the phonon DOS of the Mg$_2$Sn reservoir calculated from these two methods, as shown in Fig. 5(c). By comparing phonon DOS calculated from the direct method, the MA and HOFCM of the DFT calculations, we found that the HOFCM gives almost identical phonon DOS results as the direct method and the MA renders a much smaller gap than the HOFCM and the direct method.

In addition to the different DOS results calculated from the MA and HOFCM which can easily explain the difference in zero phonon transmission gap, the over-prediction of the phonon transmission at high-frequency ($> 0.3 \times 10^{14}$ rad/s) by the MA, as shown in Fig. 5(a), is likely due to the absence of the local force field scattering in the MA. This could be understood from the wave nature of phonons. Phonons are more likely to be scattered by the strain field whose characteristic size is comparable to the phonon wavelength $\lambda$. Since the strain field around the sharp interface spans a few unit cells, the phonons with small wavelength or large wavenumber ($k = 2\pi/\lambda$) are strongly affected. According to the phonon dispersion in Fig. 2, low-frequency phonons are the modes in LA and TA branches with small wavenumber. On the other hand, phonons away from the Γ point are of middle and high frequency. Therefore, the transmission function of low-frequency phonons calculated from the MA and the HOFCM have a closer match, while the transmission of high-frequency phonons from the HOFCM is much smaller than that from the MA due to the strain field scattering.

Due to the over-predicted phonon transmission for the entire phonon spectra, the interfacial thermal conductance is overestimated by the MA, as shown in Fig. 6. The discrepancy becomes more evident at high temperature. The interfacial thermal conductance from the MA at high temperature is almost twice as large as that from the HOFCM. At low temperature the thermal conductance is mainly contributed by low-frequency phonons and both methods give similar phonon transmission for these low-frequency phonons. At high temperature, the high-frequency phonons, whose transmission is more significantly over-estimated under the MA, are excited and begin to participate in the energy transport.

## C. Phonon transmission across Mg$_2$Si/Mg$_2$Si$_{1-x}$Sn$_x$/Mg$_2$Si

In this section, we study the phonon transmission and scattering across Mg$_2$Si/Mg$_2$Si$_{1-x}$Sn$_x$/Mg$_2$Si structures with the HOFCM to explore the phonon transport across interfaces made up of a crystal and its alloys. This kind of interfaces play an important role in increasing the figure of merit of the "nanoparticle in alloy" thermoelectric materials by reducing its thermal conductivity below the alloy limit.[54, 55] In "nanoparticle in alloy" thermoelectric materials, the short-wavelength phonons are scattered by alloy scattering while the long-and-middle-wavelength phonons are scattered by the nanoparticles whose size can be tuned. Great progress has been made over the past decade on developing more efficient thermoelectric materials using such nanostructuring approach.[3] However, the detailed and systematic understanding on frequency-dependent phonon transmission across such kind of interfaces has not been developed.

Figure 7(a) illustrates the atomic system used in the AGF simulations, where an Mg$_2$Si$_{1-x}$Sn$_x$ alloy layer with a thickness of $L$ unit cells is sandwiched between two semi-infinite Mg$_2$Si crystals. The dimension of the cross-section is $2a_0 \times 2a_0$ and the periodic boundary condition is imposed when the simulation system is relaxed. Figure 7(b) shows the phonon transmission across the alloy layer with different Sn compositions, $x$, when the length of the alloy layer $L$ is fixed at 10 unit cells. When $x$ is small, the phonon transmission always decreases with the increase of $x$ for the entire phonon spectra due to the increasing scattering events in the alloy layers. The transmission curves exhibit a similar shape, where the peaks and valleys in the curves occur at the almost the same phonon frequency when the Sn concentration ($x$) is low. By examining the phonon transmission through alloy layers with other thickness, we even found that the phonon transmission across the alloy Mg$_2$Si$_{1-x}$Sn$_x$ with a thickness of $2L$, $T_{2L}^x(\omega)$, can be well approximated by that across the alloy $Mg_2Si_{1-2x}Sn_{2x}$ with a thickness of $L$ but with $2x$ Sn concentration $T_L^{2x}(\omega)$, as shown in Fig. 8(a) and (b). This indicates that the transmission is mainly determined by the total number of Sn atoms, which serves as alloy scattering centers, but is weakly dependent on the distribution of Sn atoms. Such a scaling relation indicates that the phonons are mainly scattered by the defects in the alloy layer and the scattering events due to the Sn defects can be regarded as independent events. However, when $x > 0.4$, such simple scaling relation does not work well. $T_{2L}^{0.4}(w)$ exhibits very different features from $T_L^{0.8}(w)$, as shown in Fig. 8(c). For example, there is a zero-transmission region for middle-frequency phonons, which is the signature of phonon DOS in $Mg_2Sn$. This is understandable because when $x = 1.0$, the interface becomes a sharp $Mg_2Si/Mg_2Sn$ interface, but as $x$ decreases, the Si atoms within the alloy layers act as defects leading to additional scattering.

To observe the collective behaviors of all phonons transmitting through the structure, the thermal resistance, or the inverse of the thermal conductance, across the structures with different $L$ at 300 K, was calculated and shown in Fig. 9. The total thermal resistance is found to be proportional to the length for each Sn concentration $x$, and their relation can be well fitted by

$$R_{total} = 2R_{interface} + \frac{L}{K_{alloy}}, \tag{12}$$

In this relation, $R_{total}$ is the total thermal resistance, the coefficients $R_{interface}$ and $K_{alloy}$ can be interpreted as the interfacial resistance of each $Mg_2Si/Mg_2Si_{1-x}Sn_x$ interface and the effective thermal conductivity of the alloy layer. By linear fitting of this relation, the interfacial resistance of $Mg_2Si/Mg_2Si_{1-x}Sn_x$ interfaces, as well as the thermal conductivity of $Mg_2Si_{1-x}Sn_x$ alloy, can be extracted.

The interfacial thermal resistance can be described as a piecewise linear function with $x$, as shown in Fig. 10(a). The value of the interfacial resistance is small and weakly dependent on the composition when $x<0.4$. This trend is consistent with the observation that the phonon transmission is dominated by the individual scattering events within the alloy layer when $x$ is small. For $x>0.4$, the slope of the interfacial resistance as a function of the Sn composition abruptly jumps to a larger value and the scattering at the interface dominates the transmission process. The two regimes of interfacial thermal conductance can also be understood by looking at the transition of phonon DOS with the composition shown in Fig. 11. While the DOS of the alloys retains the shape for high-frequency phonons as that of $Mg_2Si$ at a small $x$, the peaks seem to disappear when $x$ is larger than 0.4. Therefore, high-frequency phonons from the $Mg_2Si$ phase are less likely to transmit across the $Mg_2Si/Mg_2Si_{1-x}Sn_x/Mg_2Si$ interface due to the relatively small number of phonons available to match these incoming phonons. As a result, the interfacial resistance is larger at a large $x$ because of the important role of high-frequency phonons on thermal conductance according to the Landauer formulism, Eq. (15).

Figure 10(b) shows the effective lattice thermal conductivity of $Mg_2Si_{1-x}Sn_x$ as a function of the Sn composition, extracted from the AGF calculations, in comparison with the experimental data[46] as well as recent theoretical calculations based on the PBTE theory with interatomic force constants from DFT as inputs.[22] Overall the calculated values from the AGF resemble the typical experimental trends of alloy thermal conductivity, where the thermal conductivity drops first, then becomes nearly independent of composition and increases again when $x$ increases from 0.1 to 0.9. However, the calculated value of thermal conductivity is smaller than the measured data taken from thermoelectric handbook.[46] This observation is counter-intuitive. One would expect that anharmonicity happening in the experimental samples that are not implemented in AGF calculations would lead to a smaller thermal conductivity than the calculated values. However, Paython *et al* had found the similar results in their atomic simulations several decades ago.[56] They provided the qualitative explanation as follows. Under the harmonic approximation, the impurities destroy the translational symmetry so that some of the phonon modes cannot travel from one end to the other end and become localized modes, which do not contribute the heat flux. But with the help of anharmonic coupling, an energy exchange between the localized modes is induced, thus the heat flux is enhanced. As a result, the thermal conductivity of the disorder anharmonic crystal (experimental data) would be even larger than the disordered harmonic crystal (as calculated using the AGF model).

## IV. Conclusions

In summary, we employ the HOFCM, which originates from the MA but considers the local force difference due to the different species (including both atom types and lattice constant difference), to efficiently calculate the harmonic force constants of the interfacial regions, and then integrate the obtained force constants from the first-principles calculations with the AGF approach to study phonon transmission across $Mg_2Si/Mg_2Si_{1-x}Sn_x$ interface. Starting from the harmonic interatomic force constants from the VCA, the HOFCM uses the higher-order terms which are related to the atomic displacement and atomic species to improve the accuracy of the harmonic force constants from the VCA. The HOFCM is found to be computationally affordable to extract the harmonic interatomic force constants from DFT calculations by taking the advantage of the VCA when comparing with the direct method, while yielding phonon dispersion closer to the direct DFT method in comparison with the MA. DFT-AGF approach was developed using the force constants calculated from both the MA and the HOFCM. It is found that the MA over-predicts the phonon transmission and interfacial thermal conductance across $Mg_2Si/Mg_2Sn$ interface due to the absence of the internal strain field scattering for high frequency phonons in the MA. The frequency-dependent phonon transmission across an interface between a crystal and an alloy, which often appears in high efficiency "nanoparticle in alloy" thermoelectric materials was studied. The interfacial thermal resistance across $Mg_2Si/Mg_2Si_{1-x}Sn_x$ interface is found to be weakly dependent on the composition of Sn when the composition is less than 40%, but increases rapidly when it is larger than 40% due to the transition of the high-frequency phonon DOS in $Mg_2Si_{1-x}Sn_x$ alloys.

# Appendix: Imposing translational and rotational invariance to the fitting for interatomic force constants

First-principles calculations are performed to obtain the displacement-type-force relation, Eq. (5), which are linear equations with respect to the interatomic force constants $\phi$, $\psi$, $\chi$ $J$ and $G$. To obtain physically correct interatomic force constants, the translational and rotational invariances have to be considered when solving Eq. (5) with linear fitting. The translational invariance relations, which indicate that the energy of a system is a constant value when all atoms are displaced the same amount distance along a given direction, are expressed as[34, 38]

$$\sum_{\mathbf{R'}} \phi_{\mathbf{RR'}}^{\alpha\beta} = 0, \ \forall(\mathbf{R},\alpha\beta)$$

$$\sum_{\mathbf{R''}} \psi_{\mathbf{RR'R''}}^{\alpha\beta\gamma} = 0, \ \forall(\mathbf{RR'},\alpha\beta\gamma) \qquad (A1)$$

$$\sum_{\mathbf{R'}} \chi_{\mathbf{RR'R''}}^{\alpha\beta} = 0, \ \forall(\mathbf{RR'},\alpha\beta)$$

The rotational invariances, which ensure the constant system energy when all atoms are rotated a small angle along any axis, are written as[34, 38]

$$\sum_{\mathbf{R}'} \phi_{\mathbf{RR}'}^{\alpha\beta} (\mathbf{R}')^{\gamma} \varepsilon^{\beta\gamma\nu} = 0, \quad \forall (\mathbf{R}, \alpha\beta)$$

$$\sum_{\mathbf{R}''} \psi_{\mathbf{RR}'\mathbf{R}''}^{\alpha\beta\gamma} (\mathbf{R}'')^{\delta} \varepsilon^{\gamma\delta\nu} + \phi_{\mathbf{RR}'}^{\gamma\beta} \varepsilon^{\gamma\alpha\nu} + \phi_{\mathbf{RR}'}^{\alpha\gamma} \varepsilon^{\gamma\beta\nu} = 0, \quad \forall (\mathbf{RR}', \alpha\beta\gamma) \qquad (A2)$$

$$\sum_{\mathbf{R}'} \chi_{\mathbf{RR}'\mathbf{R}''}^{\alpha\beta} (\mathbf{R}')^{\gamma} \varepsilon^{\beta\gamma\nu} + G_{\mathbf{RR}''}^{\beta} \varepsilon^{\beta\alpha\nu} = 0, \quad \forall (\mathbf{RR}'', \alpha\beta)$$

where $\varepsilon^{\alpha\beta\gamma}$ is the Levi-Civita symbol.

These constraints can be observed by the weighted least squares or Lagrange multiplier approach[57] in the linear fitting process. However, the weighted least squares method involves the choice of weighting factor, which is rather arbitrary, and the Lagrange multiplier approach is more complicate to implement. Here, we propose a simple embedded method, converting the constrained problem to the unconstrained one based on singular value decomposition (SVD).

For convenience, we convert both the displacement-type-force relation, Eq. (5), to matrix form $\mathbf{Ax} = \mathbf{b}$ and the translational and rotational invariance constraints, Eq. (A1-A2), to $\mathbf{Cx} = \mathbf{0}$, where $\mathbf{x}$ is a vector whose elements are the interatomic force constants $\phi$, $\psi$, $\chi$ $J$ and $G$ that need to be determined, $\mathbf{A}$ and $\mathbf{C}$ are simply the coefficient matrix of the linear equation sets of the displacement-type-force relation and the constraints, and $\mathbf{b}$ is a column vector. The constrained linear fitting problem is to minimize $|\mathbf{Ax} - \mathbf{b}|^2$ under the constraints $\mathbf{Cx} = \mathbf{0}$.

Suppose $\mathbf{C}$ is an $m \times n$ matrix where $m$ is the number of constraints and $n$ is the number of independent interatomic force constants, the rank of $\mathbf{C}$, $r$, must be smaller than $n$. Otherwise, the vector $\mathbf{x}$ that satisfy $\mathbf{Cx} = \mathbf{0}$ has to be $\mathbf{0}$. Through SVD, we have $\mathbf{C} = \mathbf{U\Sigma V}^T$, where $\mathbf{U}$ and $\mathbf{V}$ are $m \times m$ and $n \times n$ unitary matrices, $\Sigma = diag(\sigma_1, \sigma_2, \cdots, \sigma_r, 0, \cdots 0)$ is an $m \times n$ rectangular diagonal matrix with $r$ positive numbers, $\sigma_i$, on the diagonal. It can be shown that the last $n-r$ column vectors of $\mathbf{V}$ span the null space of $\mathbf{C}$. In other words, any $\mathbf{x}$ satisfying $\mathbf{Cx}=0$ must be the linear combination of $\{\mathbf{V}_{r+1}, \mathbf{V}_{r+2}, ..., \mathbf{V}_n\}$,

$$\mathbf{x} = y_1 \mathbf{V}_{r+1} + y_2 \mathbf{V}_{r+2} + ... + y_{n-r} \mathbf{V}_n, \tag{A3}$$

where $\mathbf{V}_i$ is the $i$th column of $\mathbf{V}$ and $y_i$ is a scalar. In matrix form, Eq. (A3) is expressed as $\mathbf{x} = \bar{\mathbf{V}}\mathbf{y}$, where $\bar{\mathbf{V}}$ is the submatrix made up of the last $n-r$ column vectors of $\mathbf{V}$ and $\mathbf{y}$ is a $(n-r) \times 1$ vector with $y_i$ as its elements. Only if we can find the $\mathbf{y}*$ which make $|\mathbf{Ax}-\mathbf{b}|^2$ minimized, $\mathbf{x}* = \bar{\mathbf{V}}\mathbf{y}*$ strictly satisfying all constraints and is the solution of the original constrained least-squares problem, To find the $y*$, we plug in $\mathbf{x} = \bar{\mathbf{V}}\mathbf{y}$ to $\mathbf{Ax}=\mathbf{b}$ and obtain $(\mathbf{A}\bar{\mathbf{V}})\mathbf{y} = \mathbf{b}$. The least-squares solution of $(\mathbf{A}\bar{\mathbf{V}})\mathbf{y} = \mathbf{b}$ is $y* = \left[(\mathbf{A}\bar{\mathbf{V}})^T (\mathbf{A}\bar{\mathbf{V}})\right]^{-1} (\mathbf{A}\bar{\mathbf{V}})'\mathbf{b}$.

Therefore, the extracted interatomic force constant vector is expressed as

$$\mathbf{x}* = \bar{\mathbf{V}}\mathbf{y}* = \bar{\mathbf{V}}\left[(\mathbf{A}\bar{\mathbf{V}})^T (\mathbf{A}\bar{\mathbf{V}})\right]^{-1} (\mathbf{A}\bar{\mathbf{V}})'\mathbf{b}. \tag{A4}$$

**Acknowledgments**: This work is supported by the NSF CAREER award (Grant No. 0846561) and AFOSR Thermal Sciences Grant (FA9550-11-1-0109). This work utilized the Janus supercomputer, which is supported by the National Science Foundation (award number CNS-0821794), the University of Colorado-Boulder, the University of Colorado-Denver, and the National Center for Atmospheric Research. The Janus supercomputer is operated by the University of Colorado-Boulder.

**References:**

1    D. G. Cahill, W. K. Ford, K. E. Goodson, G. D. Mahan, A. Majumdar, H. J. Maris, R. Merlin, and S. R. Phillpot, J. Appl. Phys. **93**, 793 (2003).

2    D. G. Cahill, et al., Appl. Phys. Rev. **1**, 011305 (2014).

3    M. S. Dresselhaus, G. Chen, M. Y. Tang, R. Yang, H. Lee, D. Wang, Z. Ren, J. P. Fleurial, and P. Gogna, Adv. Mater. **19**, 1043 (2007).

4    W. Kim, R. Wang, and A. Majumdar, Nano Today **2**, 40 (2007).

5    A. G. Evans, D. Mumm, J. Hutchinson, G. Meier, and F. Pettit, Prog. Mater Sci. **46**, 505 (2001).

6    M.-S. Jeng, R. Yang, D. Song, and G. Chen, J. Heat Transfer **130**, 042410 (2008).

7    R. Yang and G. Chen, Phys. Rev. B **69**, 195316 (2004).

8    R. Yang, G. Chen, and M. S. Dresselhaus, Nano Lett. **5**, 1111 (2005).

9    R. Yang, G. Chen, M. Laroche, and Y. Taur, J. Heat Transfer **127**, 298 (2005).

10   N. Mingo and L. Yang, Phys. Rev. B **68**, 245406 (2003).

11   J.-S. Wang, J. Wang, and J. Lü, Eur. Phys. J. B **62**, 381 (2008).

12   X. Li and R. Yang, Phys. Rev. B **86**, 054305 (2012).

13   H. Zhao and J. B. Freund, J. Appl. Phys. **105**, 013515 (2009).

14   W. Zhang, T. Fisher, and N. Mingo, J. Heat Transfer **129**, 483 (2007).

15   X. Li and R. Yang, J. Phys.: Condens. Matter **24**, 155302 (2012).

16   Z. Tian, K. Esfarjani, and G. Chen, Phys. Rev. B **86**, 235304 (2012).

17   T. Yamamoto and K. Watanabe, Phys. Rev. Lett. **96**, 255503 (2006).

18   L. Chen, Z. Huang, and S. Kumar, Appl. Phys. Lett. **103**, 123110 (2013).

19   Z. Huang, T. Fisher, and J. Murthy, J. Appl. Phys. **109**, 074305 (2011).

20   A. Ward, D. Broido, D. A. Stewart, and G. Deinzer, Phys. Rev. B **80**, 125203 (2009).

21   L. Lindsay, D. Broido, and T. Reinecke, Phys. Rev. B **87**, 165201 (2013).

22   W. Li, L. Lindsay, D. Broido, D. A. Stewart, and N. Mingo, Phys. Rev. B **86**, 174307 (2012).

23  S. Lee, K. Esfarjani, T. Luo, J. Zhou, Z. Tian, and G. Chen, Nature Commun. **5** (2014).

24  K. Esfarjani, G. Chen, and H. T. Stokes, Phys. Rev. B **84**, 085204 (2011).

25  M. N. Luckyanova, et al., Science **338**, 936 (2012).

26  J. Garg, N. Bonini, B. Kozinsky, and N. Marzari, Phys. Rev. Lett. **106**, 045901 (2011).

27  Z. Tian, J. Garg, K. Esfarjani, T. Shiga, J. Shiomi, and G. Chen, Phys. Rev. B **85**, 184303 (2012).

28  T. Murakami, T. Shiga, T. Hori, K. Esfarjani, and J. Shiomi, Europhys. Lett. **102**, 46002 (2013).

29  J. M. Larkin and A. J. McGaughey, J. Appl. Phys. **114**, 023507 (2013).

30  J. Zhou, X. Li, G. Chen, and R. Yang, Phys. Rev. B **82**, 115308 (2010).

31  V. Zaitsev, M. Fedorov, E. Gurieva, I. Eremin, P. Konstantinov, A. Y. Samunin, and M. Vedernikov, Phys. Rev. B **74**, 045207 (2006).

32  Q. Zhang, J. He, T. Zhu, S. Zhang, X. Zhao, and T. Tritt, Appl. Phys. Lett. **93**, 102109 (2008).

33  O. Dubay and G. Kresse, Phys. Rev. B **67**, 035401 (2003).

34  K. Esfarjani and H. T. Stokes, Phys. Rev. B **77**, 144112 (2008).

35  X. Tang and B. Fultz, Phys. Rev. B **84**, 054303 (2011).

36  S. Baroni, S. de Gironcoli, A. Dal Corso, and P. Giannozzi, Reviews of Modern Physics **73**, 515 (2001).

37  S. De Gironcoli, P. Giannozzi, and S. Baroni, Phys. Rev. Lett. **66**, 2116 (1991).

38  S. de Gironcoli, Phys. Rev. B **46**, 2412 (1992).

39  S. De Gironcoli, E. Molinari, R. Schorer, and G. Abstreiter, Phys. Rev. B **48**, 8959 (1993).

40  S. de Gironcoli and S. Baroni, Phys. Rev. Lett. **69**, 1959 (1992).

41  S. Baroni et al., http://www.quantum-espresso.org.

42  L. G. Rego and G. Kirczenow, Phys. Rev. Lett. **81**, 232 (1998).

43  S. Pettersson and G. Mahan, Phys. Rev. B **42**, 7386 (1990).

44  G. Chen, Appl. Phys. Lett. **82**, 991 (2003).

45  J.-i. Tani and H. Kido, Physica B: Condensed Matter **364**, 218 (2005).

46  D. M. Rowe, *Thermoelectrics handbook: macro to nano* (CRC press, 2005).

47  P. Giannozzi, et al., J. Phys.: Condens. Matter **21**, 395502 (2009).

48  J. P. Perdew and A. Zunger, Phys. Rev. B **23**, 5048 (1981).

49  E. Bitzek, P. Koskinen, F. Gähler, M. Moseler, and P. Gumbsch, Phys. Rev. Lett. **97**, 170201 (2006).

50  M. Hutchings, T. W. D. Farley, M. Hackett, W. Hayes, S. Hull, and U. Steigenberger, Solid State Ionics **28**, 1208 (1988).

51  R. Kearney, T. Worlton, and R. Schmunk, J. Phys. Chem. Solids **31**, 1085 (1970).

52  J.-i. Tani and H. Kido, Comput. Mater. Sci. **42**, 531 (2008).

53  Q. Meng, L. Wu, and Y. Zhu, Phys. Rev. B **87**, 064102 (2013).

54  S. Wang and N. Mingo, Appl. Phys. Lett. **94**, 203109 (2009).

55  N. Mingo, D. Hauser, N. Kobayashi, M. Plissonnier, and A. Shakouri, Nano Lett. **9**, 711 (2009).

56  D. N. Payton III, M. Rich, and W. M. Visscher, Phys. Rev. **160**, 706 (1967).

57  N. Mingo, D. A. Stewart, D. A. Broido, and D. Srivastava, Phys. Rev. B **77**, 033418 (2008).

**Figure Captions:**

Figure 1. (a) Calculated phonon dispersions of $Mg_2Si$ and $Mg_2Sn$ using the force constants from the first-principles calculations: the direct method, the MA, and the HOFCM, in comparison with the measured phonon dispersions from inelastic neutron scattering experiments.[30, 31] (b) Calculated phonon DOS of $Mg_2Si$ from the direct method with and without including the long-range interaction due to the Born effective charge.

Figure 2. Phonon dispersions of stretched $Mg_2Si$ and compressed $Mg_2Sn$ computed using the force constants from the direct method, the MA, the HOFCM. $Mg_2Si$ and $Mg_2Sn$ are strained to the same averaged lattice constant of them to mimic the deformed lattice near the interface.

Figure 3. (a) Lattice structure of $Mg_2Si/Mg_2Sn$ superlattice. The small, medium and large atoms are Mg, Si and Sn, respectively. (b) Phonon DOS of $Mg_2Si/Mg_2Sn$ superlattice calculated using the force constants from the direct method, the MA, the HOFCM.

Figure 4. (a) The atomic system used to perform energy minimization. The arrows indicate the periodic boundary condition. (b) The atomic system for the AGF calculation. The yellow, cyan and pink atoms are Si, Sn and Mg, respectively.

Figure 5. (a) Frequency-dependent phonon transmission across a sharp $Mg_2Si/Mg_2Sn$ interface. (b) Phonon DOS of $Mg_2Si$. (c) Phonon DOS of $Mg_2Sn$.

Figure 6. Interfacial thermal conductance across $Mg_2Si/Mg_2Sn$ interface at different temperature. The MA overpredicts the interfacial thermal conductance.

Figure 7. (a) The schematic of the simulation system of $Mg_2Si/Mg_2Si_{1-x}Sn_x/Mg_2Si$ structure. (b) Phonon transmission across the $Mg_2Si/Mg_2Si_{1-x}Sn_x/Mg_2Si$ structure as a function of phonon frequency for different Sn compositions (x). The length of the alloy layers is fixed at 10 unit cells.

Figure 8. Phonon transmission across the $Mg_2Si/Mg_2Si_{1-x}Sn_x/Mg_2Si$ structure as a function of phonon frequency for alloy layers with different Sn compositions and thickness.

Figure 9. The total thermal resistance of $Mg_2Si/Mg_2Si_{1-x}Sn_x/Mg_2Si$ structure as a function of the thickness of alloy layers.

Figure 10. (a) The extracted interfacial resistance of $Mg_2Si/Mg_2Si_{1-x}Sn_x$ interface as a function of x. (b) The extracted effective thermal conductivity of $Mg_2Si_{1-x}Sn_x$ alloy as a function of x, in comparison with the measured data from Ref. [46] and the PBTE calculation from Ref. [22].

Figure 11. The calculated phonon DOS of $Mg_2Si_{1-x}Sn_x$ alloys.

**Figures:**

**Figure 1**

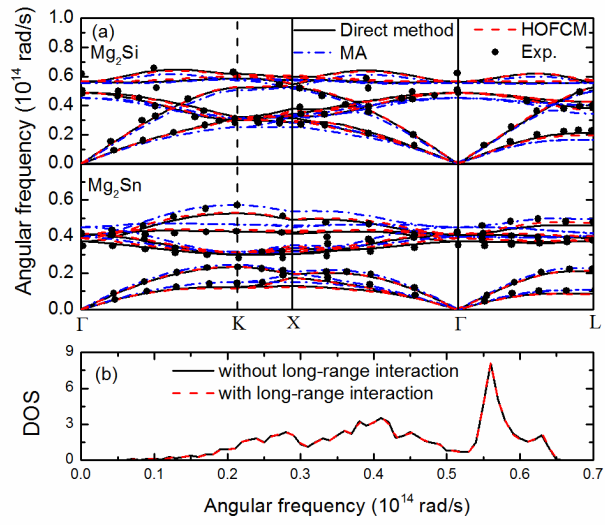

**Figure 2**

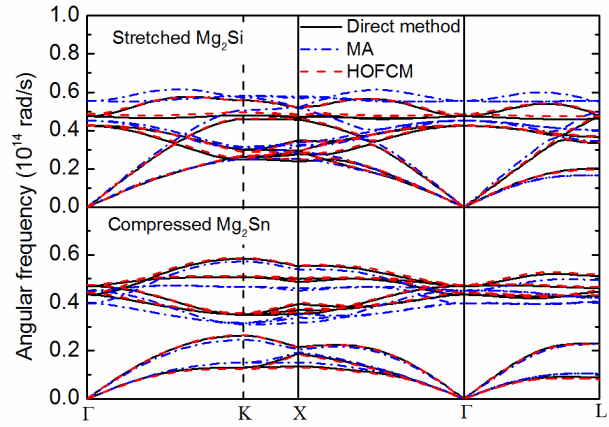

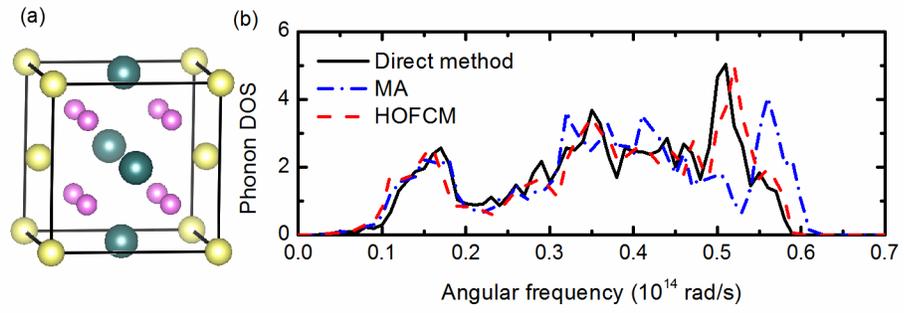

**Figure 4**

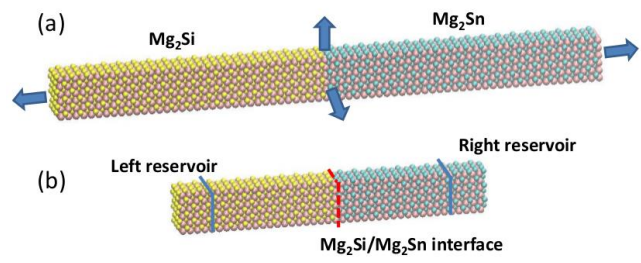

**Figure 5**

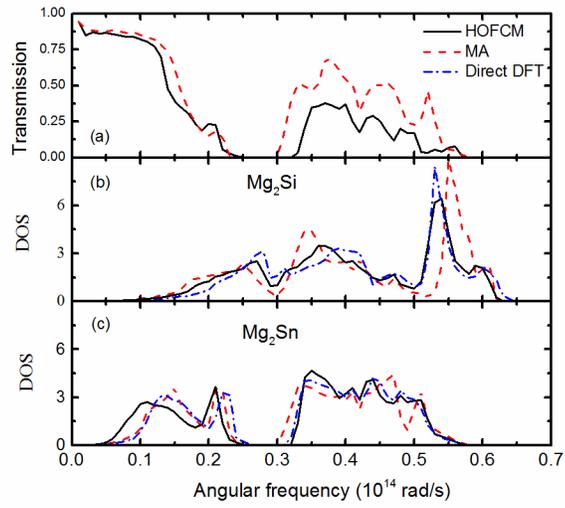

**Figure 6**

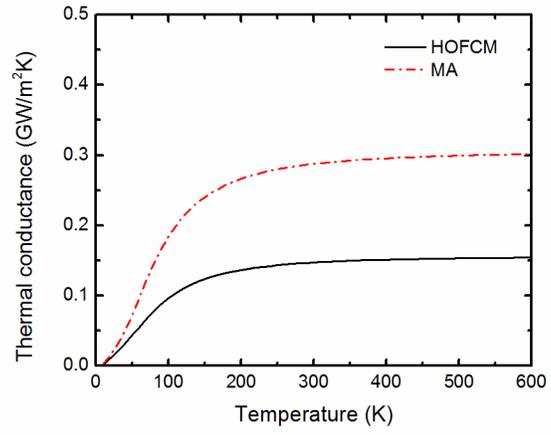

**Figure 7**

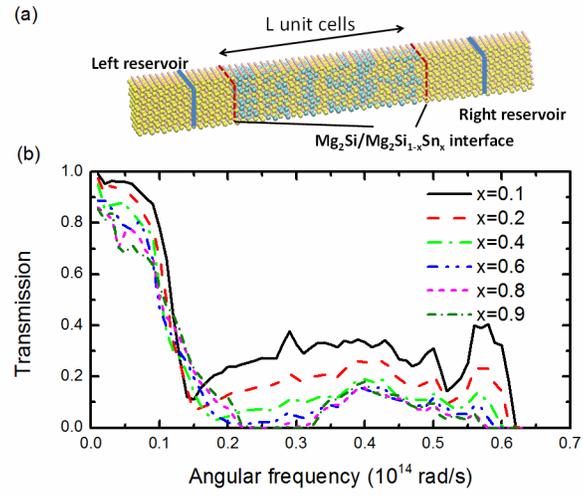

**Figure 8**

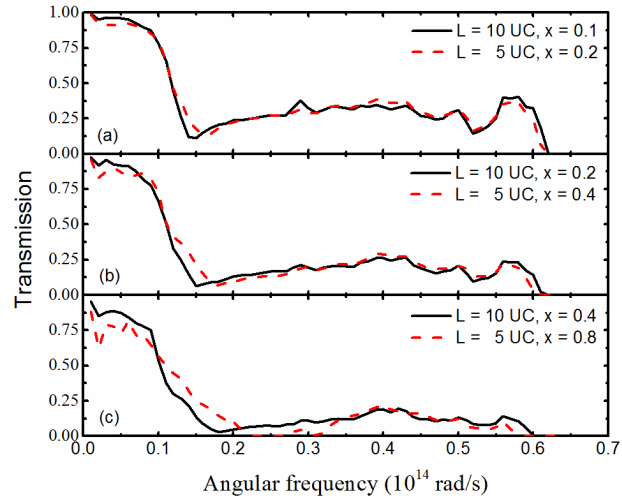

**Figure 9**

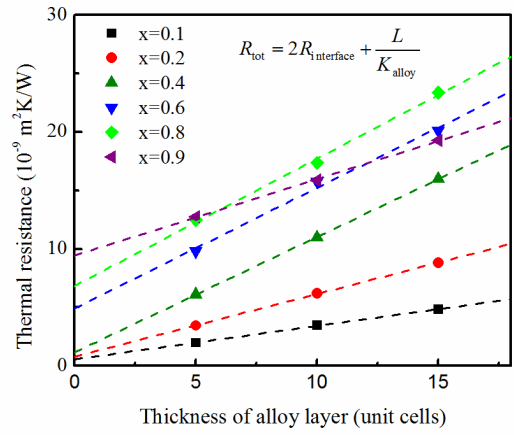

**Figure 10**

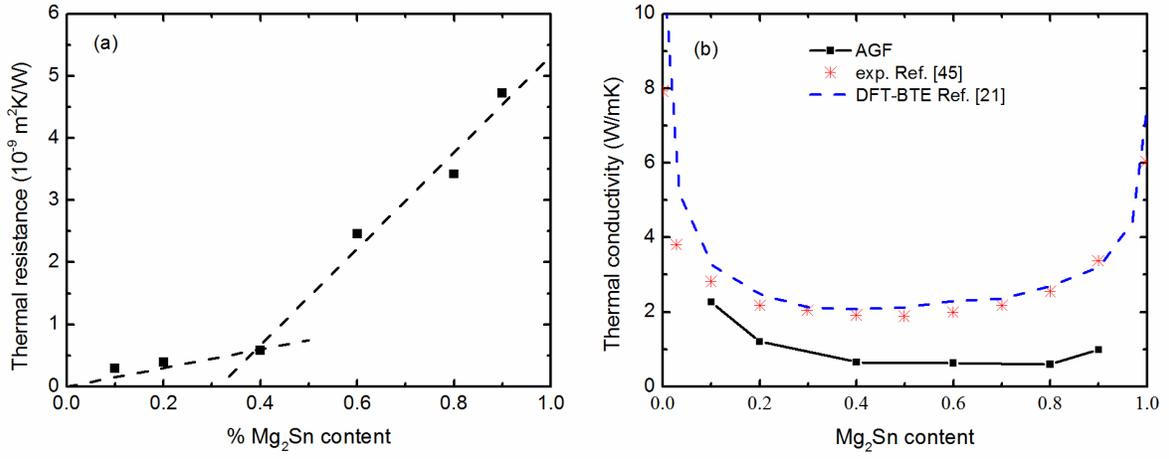

**Figure 11**

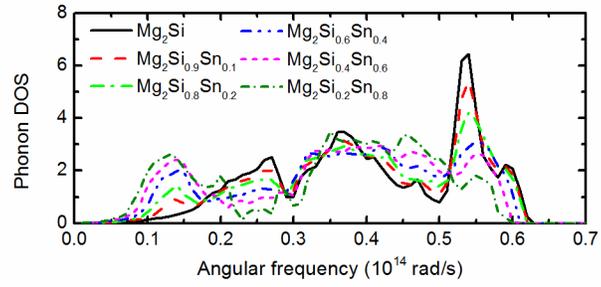